# A Multi-Agent Generative AI Framework for IC Module-Level Verification Automation

Wenbo Liu, wenbo.liu@jaguarmicro.com

Forbes Hou, forbes.hou@jaguarmicro.com

Jon Zhang, jon.zhang@jaguarmicro.com

Hong Liu, hong.liu@jaguarmicro.com

Allen Lei, allen.lei@jaguarmicro.com

IC Design Department, Jaguar Microsystems, China

*Abstract*—As large language models demonstrate enormous potential in the field of Electronic Design Automation (EDA), generative AI-assisted chip design is attracting widespread attention from academia and industry. Although these technologies have made preliminary progress in tasks such as code generation, their application in chip verification - a critical bottleneck in the chip development cycle - remains at an exploratory stage. This paper proposes an innovative Multi-Agent Verification Framework (MAVF) aimed at addressing the limitations of current single-LLM approaches in complex verification tasks. Our framework builds an automated transformation system from design specifications to testbench through the collaborative work of multiple specialized agents, including specification parsing, verification strategy generation, and code implementation. Through verification experiments on multiple chip modules of varying complexity, results show that MAVF significantly outperforms traditional manual methods and single-dialogue generative AI approaches in verification document parsing and generation, as well as automated testbench generation. This research opens new directions for exploring generative AI applications in verification automation, potentially providing effective approaches to solving the most challenging bottleneck issues in chip design.

## I. INTRODUCTION

With the increasing complexity of integrated circuit design, verification workload has grown dramatically, becoming a major bottleneck in chip development cycles. Although the industry has mature solutions for testbench framework generation and regression testing automation, the automation level in specification comprehension and verification development stages remains insufficient, heavily relying on verification engineers' professional experience and manual judgment.

The automation of current verification processes faces two major challenges:
1. Challenges in Specification Understanding and Documentation Transformation:
    a) Design specifications contain non-standardized multimodal information (textual descriptions, diagrams, etc.)
    b) Verification engineers need to manually extract key information and transform it into verification plans
    c) Information extraction and transformation processes are time-consuming and error-prone
    d) Low efficiency in writing verification-related documents (verification plans, testbench specifications, etc.)
2. Automation Bottlenecks in Core Verification Environment Development:
    a) Core components such as interface drivers, monitors, and reference models still heavily depend on manual development
    b) Test scenario case design and implementation require extensive professional experience
    c) Existing tools cannot effectively support the automated development of these core components and test cases

Recently, Generative AI technology has achieved significant breakthroughs in document analysis and code generation[13][16][17]. However, due to the deep domain expertise required in chip verification, simple dialogue-based approaches struggle to solve such complex problems[1][18][19]. Notably, emerging technologies such as Retrieval-Augmented Generation (RAG), Agents, and Multi-agent systems have matured, offering new possibilities for solving verification automation challenges. These technologies can:
- Handle complex tasks through collaboration of multiple specialized agents
- Better understand and process domain-specific knowledge
- Provide more reliable automation solutions for verification processes

Based on this background, this paper proposes introducing multi-agent technology into module-level chip verification processes, aiming to improve verification efficiency, reduce manual dependencies, and achieve intelligent transformation of verification processes.

TABLE I
Comparison between Traditional Verification Methods and Intelligent Verification Methods

| Phase | Traditional Approach | Intelligent Approach |
|---|---|---|
| Specification Analysis | Manual Analysis | Auto-generation + Manual Calibration |
| Verification Planning | Manual Analysis | Auto-generation + Manual Calibration |
| Environment Setup | Framework Auto-generation + Manual Coding | Auto-generation + Manual Calibration |

Since module-level verification typically consumes a significant portion of verification team resources in chip development projects, and is relatively independent compared to other levels of verification work, this paper primarily focuses on how to utilize Multi-Agent technology to accelerate the early stages of module-level verification processes.

The main contributions of this paper can be summarized in three aspects:
1. Theoretical Innovation
    a) Proposes a systematic approach to applying multi-agent collaborative frameworks in chip verification
    b) Innovatively decomposes complex verification processes into collaboration between subtasks and establishes inter-agent coordination mechanisms
    c) Designs verification domain-specific agent role definitions and interaction protocols
2. Technical Breakthroughs
    a) Implements standardized representation methods for verification specifications, supporting precise information transfer between agents
    b) The proposed framework effectively addresses the limitations of general large models in complex chip verification tasks through specialized agent collaboration
3. Practical Value
    a) Applications in multiple actual chip verification projects prove the framework's effectiveness in improving verification efficiency
    b) Provides a scalable verification automation solution that easily integrates into existing verification processes
    c) Evaluates framework performance under different model selections

Based on this framework, verification engineers' workload can be reduced in extracting effective information from less standardized design specification documents, and it optionally enables automated generation of complete testbench code or incremental code generation based on existing testbench.

Using the framework proposed in this paper, across different module-level designs, the accuracy of correctly generating documents and code improved from 13% to 70% compared to simple dialogue-based approaches, reducing human effort by 83%, 73%, and 50% in simple, moderate, and complex module-level verification work, respectively.

## II. RELATED WORK

*A. IC Verification Automation*
- Traditional Verification Methodology

In traditional chip module-level verification processes, verification engineers need to perform the following work sequentially:
1. Interpret design specification documents, extract key information and formulate verification plans
2. Develop testbench components (interface drivers, monitors, reference models, etc.)
3. Design and implement test scenarios
4. Execute regression testing and analyze results
5. Evaluate verification completeness

- Existing Automation Tools and Frameworks
    - Industry typically builds testbenches based on UVM methodology for module-level functional verification. Due to the high commonality in framework, hierarchy, and components of testbench for different DUTs under UVM methodology, companies usually build internal testbench generation tools that align more closely with their internal processes. These tools can easily generate basic frameworks including testbench file hierarchy, Agent framework, reference models, and scoreboard components. On top of this framework, specific interface stimuli and core logic of

reference models must be manually implemented according to specific DUT specifications and verification requirements[15].
- Advanced EDA vendors provide a series of tools for test point management and regression management. Based on these tools, iterations of simulation execution, debug modification, and re-simulation can be performed on completed testbench, ultimately achieving coverage-driven verification convergence through numerous iterations.
- Current Limitations
  1. Despite significant improvements in verification process automation by existing EDA tools, critical technical bottlenecks remain. First, in the design specification parsing and verification document generation phase, design specifications typically present non-standardized multimodal information, including timing constraints and functional definitions in natural language, as well as waveforms and state transition diagrams. This heterogeneous information integration poses dual challenges for verification engineers: manually identifying potential contradictions between different document versions and transforming fragmented information into standardized descriptions for verification needs. While existing tools provide document templates, they lack deep semantic parsing capabilities, leaving core tasks like test point extraction and scenario coverage analysis dependent on manual experience.
  2. In the testbench construction phase, current automation technology is mainly limited to framework code generation. While mainstream tools can create UVM testbench directory structures and basic class frameworks, significant limitations exist in developing core functional components. Specifically, key aspects such as interface protocol driver timing control logic, design behavior-based prediction model construction, and intelligent generation of coverage convergence strategies still require engineers to manually write substantial custom code. This limitation directly leads to efficiency losses in both technical and methodological aspects: verification component development requires repeated implementation of common design patterns, and test scenario construction heavily depends on engineers' secondary interpretation of design specifications.

*B. Fundamental Technologies*

Modern artificial intelligence breakthroughs provide new technical approaches for solving chip verification automation challenges[18][19]. This framework primarily relies on three core technical systems: large language models' semantic understanding capabilities, retrieval-augmented generation's knowledge integration mechanisms, and multi-agent systems' collaborative work paradigms.

Large language models, as core carriers of cognitive computing, demonstrate deep parsing capabilities for multimodal design specifications. Their semiconductor domain knowledge graph, obtained through pre-training, can effectively identify key design elements such as timing constraints and state transition logic in natural language descriptions.

Retrieval-Augmented Generation (RAG) significantly improves technical solution reliability through a dual-layer architecture of domain knowledge base and verification experience base. The domain knowledge base integrates structured knowledge like design-related specifications and interface protocol standards, while the verification experience base accumulates implicit experience from historical projects, including test scenario patterns and coverage convergence strategies.

The multi-agent framework achieves procedural decomposition of complex verification tasks through specialized division and collaboration mechanisms[2]. Agents form a closed-loop workflow under unified coordination: design parsing agents extract key design features, verification planning agents formulate test strategies, code generation agents implement components, and quality assurance agents perform consistency checks.

*C. Generative AI Application Opportunities in IC Verification*

The potential of AI technology in module-level verification manifests in three interconnected dimensions: semantic parsing of design specifications, intelligent construction of verification plans, and automated generation of verification components[3][4].

In design specification processing, multimodal understanding systems can establish semantic association networks between design elements by integrating natural language processing and graphic symbol recognition technologies. For example, the system can automatically identify interface topology relationships in design diagrams while converting natural language descriptions in design specifications into structured representations.

In testbench construction, the multi-agent collaborative mechanism enables specialized division of tasks, with agents focusing on interface protocol conversion, design behavior modeling, and stimulus combination generation. This distributed architecture not only improves code generation efficiency but also enhances functional implementation accuracy through cross-validation mechanisms.

Notably, the deep application of AI technology is reshaping the verification methodology architecture. Traditionally discrete verification document writing, testbench development, and test execution phases are now organically connected through intelligent systems' semantic understanding and knowledge reasoning capabilities.

This connection enables verification processes to perform reverse optimization based on coverage analysis results, such as automatically identifying uncovered scenarios and retroactively correcting verification plans.

## III. FRAMEWORK IMPLEMENTATION

### A. Multi-Agent technology

Agent is an autonomous decision-making system based on Large Language Models (LLMs), featuring goal-oriented and iterative execution capabilities[6][14]. Unlike traditional question-answering LLM applications (such as ChatGPT), agent systems possess autonomous planning and continuous execution abilities without requiring human intervention at each interaction step. In LLM-based autonomous agent architectures, LLM serves as the core reasoning engine, complemented by the following key functional modules:

- Task Planning System
    a) Goal Decomposition Mechanism: Employs hierarchical task decomposition strategies to break down complex tasks into executable atomic task units.
    b) Adaptive Optimization: Achieves continuous optimization of task execution through feedback loops, including error detection, result evaluation, and strategy adjustment.
- Memory Management System
    a) Working Memory: Maintains immediate contextual information during task execution.
    b) Long-term Storage: Implements persistent storage and efficient retrieval of large-scale knowledge based on vector databases.
- Tool Integration System
    a) Supports external API calls to extend model capability boundaries, enabling data acquisition, code execution, and other functions.
    b) Provides proprietary data source access interfaces to supplement model knowledge gaps in specific domains.

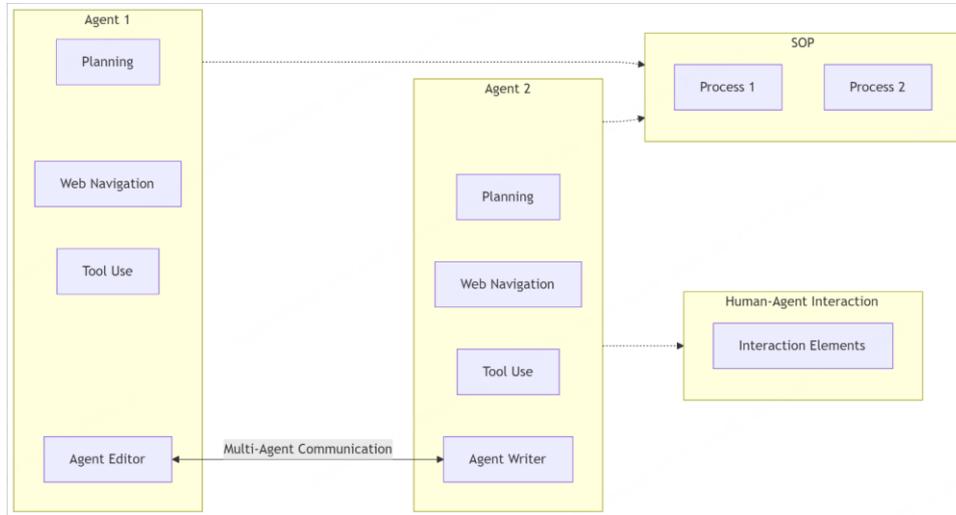

Figure 1. Multi-Agent Framework.

Multi-agent systems are distributed intelligence frameworks based on collaborative working mechanisms. By encoding Standard Operating Procedures (SOP) into structured prompts, they provide systematic solutions to complex problems[2]. This framework requires agents to participate in collaboration as domain experts, producing standardized documents including requirement specifications, system architecture designs, and workflows. These structured outputs manifest as enhanced Chain-of-Thought reasoning at the single agent level, while forming semantically clear and goal-oriented context information at the system level. Through precise role division and task decomposition mechanisms, the output quality of Large Language Models (LLMs) is significantly improved[7][11][12].

- Core Features
    a) Solution Stability: Addressing the limitations of general large models in domain-specific knowledge applications, this framework achieves output consistency and accuracy by systematically integrating industry Standard Operating Procedures (SOP) into the multi-agent system.
    b) Role Assignment Diversity: Ensures systematic and comprehensive problem-solving processes by assigning differentiated professional roles to large language models.
- System Components
    a) Agents: A distributed system built on basic agent architecture, comprising multiple agents with independent LLM engines, perception mechanisms, decision systems, execution modules, and memory units working collaboratively.

- b) Environment: Provides a unified platform for agent interaction and information exchange, supporting observation data acquisition and action result distribution.
- c) Standard Operating Procedures (SOP): Standardized agent behavior control and interaction protocols to ensure orderly and efficient system operation.
- d) Quality Review: Addresses model hallucination issues through strict review mechanisms, employing multi-level cross-validation to ensure output quality, with experiments showing a 90% reduction in overall error rates.
- e) Communication Mechanism (Routing): Information transmission and interaction protocols between agents, supporting internal system collaboration, negotiation, and competition mechanisms.
- f) Event Subscription: System-level change perception mechanism ensuring all relevant agents can promptly respond, assess impacts, and make appropriate adjustments.

B. *Multi-Agent Adaptation Analysis for Module-Level Verification Process*
  1) Process Decoupling Methodology

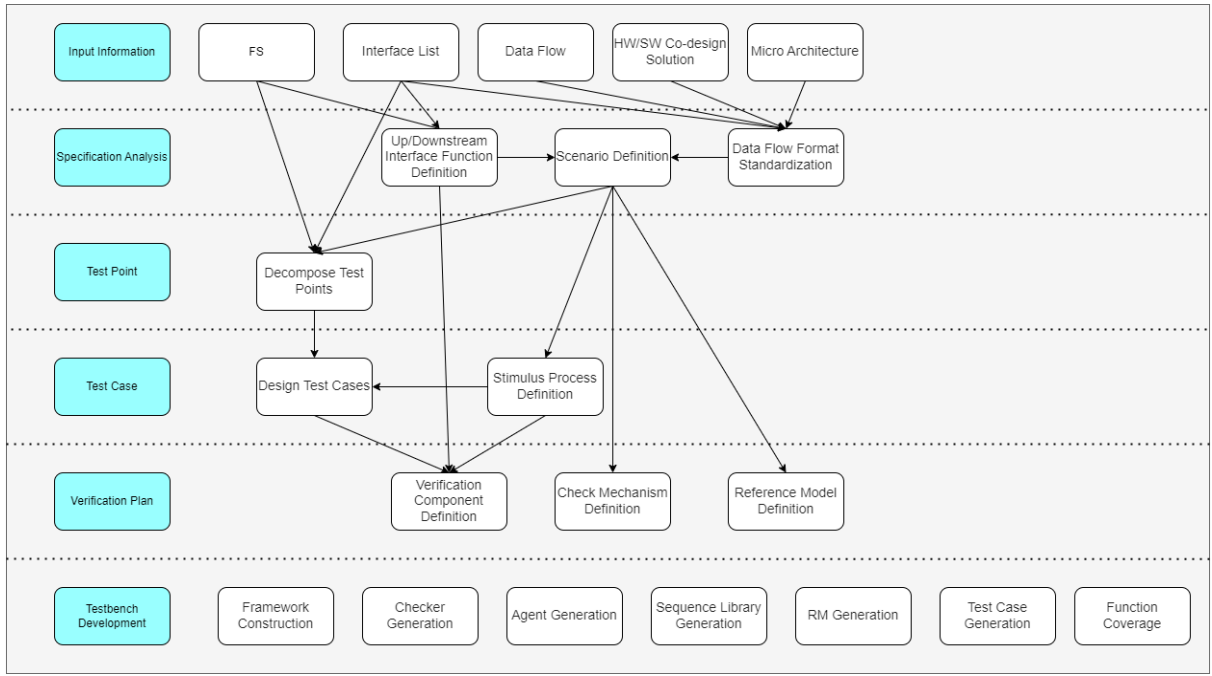

Figure 2. Module-Level Verification Workflow

The above figure shows the typical workflow of module-level verification. We decompose this workflow into the following phases:
1. Specification Analysis Phase - Accurately identify functional specifications, interface signal lists, register lists, other data structures, data flows, and working scenario configuration processes from relevant documents including design specification documents, hardware-software co-design solutions, and microarchitecture documents
2. Verification Documentation Phase - Based on the identified functional specifications, scenarios, registers, and interfaces, decompose test points of the design under test, while designing verification plans to outline implementation details of the testbench
3. Testbench Coding Phase - According to the established plans and test points, develop a comprehensive testbench including stimuli, reference models, various Agents, and test cases

These three phases are executed sequentially as the main SOP (Standard Operating Procedure) for module-level verification work. As these three phases are completed in order, various verification documents including test points and verification plans are gradually generated, and the testbench is developed according to the requirements specified in the verification documents.

Within each phase, there are different specific tasks. For example, in the specification analysis phase, various information such as functional specifications, interface lists, and data flows need to be extracted progressively from various design documents. This process serves as a sub-SOP of the specification analysis phase, and when this sub-SOP is completed, the task objectives of the specification analysis phase are achieved.

  2) Feasibility Analysis of Agent Technology

The phased characteristics of module-level verification processes provide an ideal adaptation scenario for multi-agent collaboration. Verification work can be decoupled into three phases: specification analysis, documentation generation, and testbench development, with deliverables from each phase (such as interface lists, test matrices, component code, etc.) forming strict dependencies. This atomized task structure highly aligns with the distributed processing paradigm of agent systems: each subtask (such as protocol parsing, test scenario construction, etc.) has clear input/output boundaries, and inter-agent data flow can be precisely defined.

From a technical feasibility perspective, the current evolution of large language models effectively matches verification task requirements. The 128K input window of mainstream models (such as Claude-3.5-Sonnet, DeepSeek-R1) can fully accommodate typical design specifications (approximately 5-20k tokens), while the document and code generation strategy of hierarchical phase sub-SOPs controls single outputs within 8k tokens, ensuring generation quality.

Considering the complexity of verification processes, multi-agent collaboration mechanisms demonstrate unique advantages in handling complex scenarios. When facing multi-protocol interaction scenarios, the system builds agent collaboration networks (drivers + checkers), ensures domain focus through information isolation mechanisms, and achieves real-time synchronization of changes based on a publish-subscribe pattern.

## C. Hierarchical Architecture Design

### 1) System-Level Architecture

The multi-agent verification framework achieves automated conversion from design specifications to testbench, focusing on specification comprehension and testbench setup stages in module-level verification.

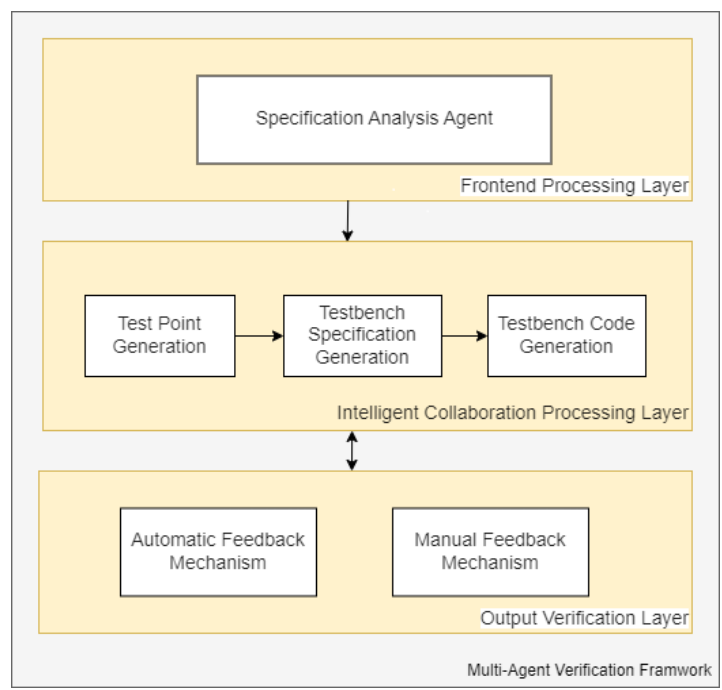

Figure 3. Multi-Agent verification framework

- Frontend Processing Layer: Unified Representation of Multimodal Input

    The Frontend Processing Layer serves as a unified entry point for multimodal inputs, with its core functionality being the transformation of heterogeneous design documents into structured design specification information. It primarily includes specification analysis agents and storage structures. The specification analysis agents generate unified format design specification information according to predefined sub-SOPs based on all existing design documents, storing the generated information for use by subsequent collaborative processing layer agents.

- Agent Collaboration Layer: Workflow Engine
    1. Employs a workflow engine to achieve dynamic scheduling and coordinated control of multiple agents. This layer defines three agents: verification plan generation agent, testbench specification generation agent, and testbench code generation agent. These different agents collaboratively execute the main SOP of the verification process, with each agent handling sub-SOPs belonging to different stages of the BT verification process. Each agent implements various capabilities required for executing its stage's sub-SOP, establishing strict work boundaries and steps between different agents while implementing specific workflows.

2. Different agents have distinct roles and responsibilities: Verification Plan Generation Agent - generates specific test point lists; Testbench Specification Generation Agent - generates testbench file hierarchy, testbench topology, Agent component detailed design, RM/Checker component detailed design, and functional coverage definitions; Testbench Code Generation Agent - completely implements various parts of the testbench according to existing testbench specifications.
- Output Verification Layer: Closed-loop Feedback Mechanism
    1. Each agent integrates a planning-execution-verification closed-loop thinking chain, implementing task decomposition and self-correction through the ReAct mode. The workflow engine manages task dependencies, automatically triggering subtasks when parent task states meet preset acceptance criteria. For example, when the verification plan generation agent completes the orthogonal coverage matrix, the system automatically awakens the testbench specification generation agent.
    2. This feedback mechanism includes both automatic feedback (such as completeness checks of design specifications, orthogonal checks between test points and FS, interface consistency checks between UVM testbench components) and allows manual intervention at key points to provide explicit modification suggestions until human feedback meets standards.

*2) Agent Collaboration Paradigm*

In the design of agent collaboration mechanisms, master-slave control is used both at the main SOP level between stages and between agents in the collaboration layer, with a separate control unit acting as a central coordinator responsible for task scheduling and resource allocation. Meanwhile, different layers and agents share necessary information such as design specifications, but process information is selectively passed between agents to minimize irrelevant background information for Generative AI operation.

Inter-agent communication protocols adopt a layered message design: the control plane transmits task state machine information (including task ID, priority, deadlines), while the data plane transmits structured task payloads (project-related background information and task-related example templates in JSON format). At key checkpoints (such as meeting test point orthogonal coverage requirements), the system generates visualization dashboards allowing engineers to intervene in execution details or inject domain knowledge.

This architecture achieves robust output through a "generate-verify-correct" iteration mechanism. Each agent must pass cross-checking by paired verification agents before proceeding to the next phase after completing staged outputs. If implementation errors are detected, the system automatically generates correction suggestions and feeds them back to the code generation phase.

*D. Agent Implementation*

Each agent primarily executes the sub-SOP workflows for the aforementioned stages. This section will introduce the work content of each agent and the key ideas and methods during execution. It should be noted that unless explicitly stated that human intervention is required, all processes are automatically executed using a Python-based software framework according to the generative AI responses and preset procedures.

*1) Specification Parsing Agent*

Design specifications typically include multiple documents in different formats and forms, including:
- FS (Functional Specification)
- Microarchitecture
- Hardware-software interaction documents
- Register lists
- Software programming manuals

The specification analysis agent implements multimodal information extraction and standardization through a multi-stage processing workflow. It first standardizes input document formats, uses multimodal generative AI models to decompose non-textual diagrams, and provides contextual descriptions for figures and tables. Then, it uses generative AI to extract key design specification information according to predetermined dimensions.

Finally, all design specifications are consolidated into the following standardized key information:
- Original FS list
- Hardware interfaces - including upstream/downstream modules and detailed functional explanations of interface signals, with special attention to custom timing interfaces
- Protocol interfaces - higher-level data structures defined above hardware interfaces for inter-module interaction
- Register information
- Working scenarios
- Per-scenario configuration processes
- Per-scenario data flows
- Complete design specifications

All information is saved in JSON format to standardize generative AI output and enhance its information parsing capabilities. The final output is a JSON collection generated according to a unified template that contains all design specification information.

*2) Verification Plan Generation Agent*

The verification plan generation agent's main task is to generate detailed test point decomposition based on the standardized design specification output from the specification parsing agent, referring to predetermined test point decomposition templates and rules. The agent then plans specific test cases with appropriate granularity. Each test case definition includes its relationship with corresponding test points, stimulus scenarios, checking mechanisms, and pass conditions.

In this framework, test points are primarily categorized along the following dimensions:
- Clock and reset
- Functional specifications
- Application scenarios
- Interface buses
- Registers
- Exceptions
- DFX
- Processing flows
- Performance

Pre-defining reasonable test point decomposition dimensions will greatly improve the accuracy and completeness of test point decomposition. Additionally, providing excellent test point decomposition examples enhances this process. Test case definitions and descriptions also require corresponding templates and examples. The examples mentioned in this paper use test point and case standards from simple modules in previous projects, but do not include design specification information for these modules.

Test case descriptions are divided into the following categories:
- Register test cases
- Basic functionality test cases
- Single-scenario random test cases
- Mixed random test cases
- Exception test cases
- Other special scenario directed test cases

After completing test case generation, a cross-reference matrix of test cases and test points serves as an evaluation standard. This matrix ensures that each test point maps to at least one test case. As a result, when all test cases are reliably executed, verification coverage for all test points is guaranteed.

*3) Testbench Specification Agent*

Based on previously identified design interfaces, registers, upstream/downstream modules, and decomposed test points and test cases, reasonable testbench specifications need to be established to guide subsequent testbench development work. This is necessary because the conversion from test points and design specifications to concrete testbench code requires a testbench solution refinement process, which better aligns with human engineers' workflow.

Following current industry practices, this framework plans the testbench based on UVM. First, the overall testbench topology needs to be established and a topological diagram should be created[5], including:
- base_test/tc_lib - Parent class for all test cases and tc_lib containing individual test cases
- top_tb - DUT instantiation
- env/checker/fcov/rm/virtual_sequencer/reg_ral - Testbench components
- Agent - Third-party or self-developed VIPs integrated into the testbench based on DUT interfaces and verification requirements
- virtual_seq/seq_lib/cfg_seq - Based on Agent
- Interface - Logic implementing connections between DUT and various Agent interfaces

Generative AI needs to determine the functionality, quantity, and hierarchical relationships of these verification components based on previous agents' output, and provide specific definitions of core data structures and driving functions within components. The key is to design collaboration methods among testbench components based on actual stimulus processes and checking mechanisms. This stage should not provide specific code but qualitatively clarify each component's specific functions. Appropriate prompts are needed to guide this stage, such as:
- Agents should include detailed functional descriptions, specific interface signals, usable VIPs, and integration examples
- Checking mechanisms
- Check objects
- End-to-end verification when possible, considering testbench complexity
- Checks at higher levels when possible

- How to predict expected DUT responses
- How to obtain actual DUT responses
- Specific forms of register model integration
- How to define functional coverage

It should be noted that during this work's implementation, human intervention is crucial at the testbench specification stage. Through multiple rounds of prompts and corrections to confirm that current testbench design specifications meet various scenario verification requirements, the accuracy of subsequent testbench code generation agent output is significantly improved. The overall modification work through human intervention at this stage is also significantly less than the workload of manual optimization after completing the fully automated process.

*4) Testbench Code Generation Agent*

In the testbench implementation stage, when verification engineers manually perform this verification process, they have thoroughly understood the design specifications and established rigorous verification plans and testbench design solutions. Therefore, in the next stage, the testbench generation agent only needs to accurately and completely implement the code writing work for the full testbench by strictly following the verification documents generated in the above steps.

The work at this stage is mainly divided into framework-level, component-level, and scenario-level parts:

- Framework Level (UVM Structure)
    a) Directory file structure generation
    b) Component framework code generation
    c) Component connection relationship code generation
    d) DUT and Agent interface connection relationship generation
- Component Level (Driver/Monitor, etc.)
    a) All item generation
    b) Item-based Driver logic and Monitor logic
    c) RM core code generation
    d) Checker core code generation
- Scenario Level (Sequence/Testcase)
    a) Generate seq lib with sanity sequences using Agent and its items to apply stimulus
    b) Generate various tests by calling seq lib

It should be specifically noted that the code framework generation technology commonly used in the industry today is completely consistent with the framework-level work described in this section. The difference is that the input source for traditional automated testbench generation technology is manually filled in by verification engineers based on their understanding of design specifications, while the input for the framework-level stage described in this paper completely depends on the previous steps of the framework described in this paper. To inherit high-value work from existing verification processes and better accommodate existing workflows, existing automation scripts can be integrated at this stage. During the implementation of this work, it was found that using existing automation scripts with fixed testbench topology and file system structure can significantly improve the stability of this process. By fixing constraints on testbench topology level to constrain subsequent output forms, it effectively addresses the instability issues of large model outputs.

*E. Key Technical Implementation*

In implementing this framework, we drew inspiration from multi-agent frameworks like MetaGPT and CrewAI. For inter-agent communication, we employed a full broadcast method for unified standardized design specifications, while utilizing a simple publish-subscribe mechanism for other agent collaborations to avoid redundant transmission of ineffective information[2][8][9].

Furthermore, we implemented the module-level chip verification multi-agent framework, building upon existing frameworks. During implementation, we experimented with various models including openai/4o-mini, anthropic/claude-3.5-sonnet, and deepseek/deepseek-r1, utilizing OpenAI-compatible interfaces for large model interactions.

In the input processing phase, we employed RAG-based incremental retrieval of original documents, carrying this information through specification analysis and verification plan generation stages. This approach avoided the need to carry the entire original design documentation through each step of the framework, thereby increasing the effective information density in the context provided to generative AI.

Additionally, we implemented both automated and manual review mechanisms with different standards at key process points to jointly enhance quality and minimize error propagation. For high-value information such as registers, Data Structures, and interfaces, we adopted JSON format for standardized storage and transmission, thereby improving the accuracy of generative AI processing.

IV. QUALITY ASSURANCE MECHANISM

From a chip verification quality-first perspective, each stage requires reliable mechanisms to ensure output quality. In traditional manual workflows, engineers conduct multiple personal checks at different nodes, while milestone reviews are performed by expert committees who thoroughly examine key deliverables and track issue resolution.

From the perspective of applying a multi-agent verification framework, its main value lies in improving verification engineers' efficiency in handling routine tasks, thereby freeing up human resources for necessary inspections. The essence here is a paradigm shift from "human generation and human inspection" to "AI generation and human inspection." Through rapid iterations between automated execution, automated review, and human review, engineers can complete individual work quickly and with high quality[10]. Meanwhile, trustworthy AI self-checking mechanisms need to be established to improve the reliability of AI-generated content, reducing the workload of human inspection.

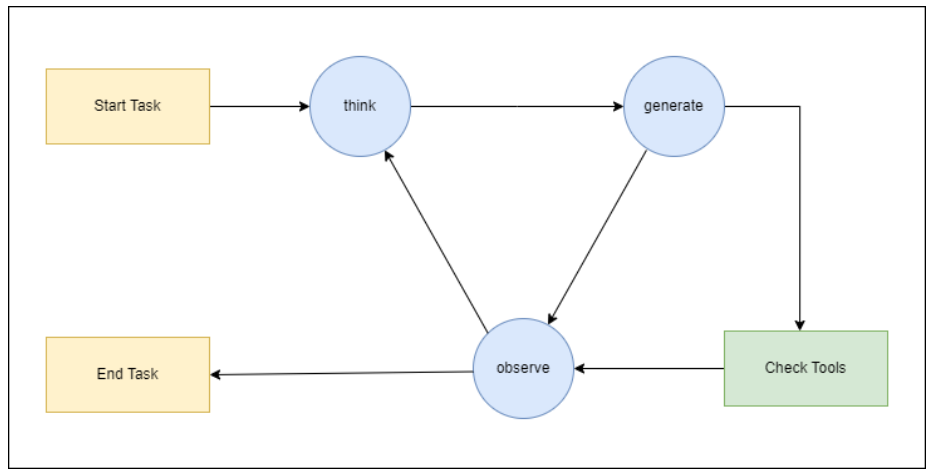

Figure 4. Dynamic Verification Loop

As shown in the dynamic verification loop diagram above, each task undergoes inspection after the initial GenAI-based forward generation, with inspection results fed into a perception process to determine if the generation meets requirements. If requirements are not met, the dynamic verification loop continues; if met, the loop exits.

Based on this dynamic verification concept, the framework implements multi-dimensional consistency checks after key work stages to ensure accuracy and completeness:
1. Specification Parsing Agent - Design Element Coverage Analysis
    a) Interfaces/Registers/Data Structures
    b) Scenarios/Configuration Flows/Exception Handling/DFX
2. Verification Planning Agent - Orthogonal Coverage Matrix Generation
    a) Analyze whether all design specifications have test point coverage
    b) Analyze which stimuli in the testbench cover each test point
3. Testbench Specification Generation Agent - Manual Review
    a) Confirm Agent planning rationality
    b) Confirm checking mechanism rationality
4. Testbench Code Generation Agent - Dual Syntax/Semantic Check
    a) Manual Review
    b) Execute smoke test cases, manual waveform inspection
5. Agent provides analysis results for human confirmation. Humans provide review feedback at critical points, forming a verification loop with human participation

From the consistency checking methods at each stage above, it's evident that complex problem execution still requires in-depth review by human engineers. It's important to note that this framework's intention is not to completely replace human engineers with automation tools. Rather, due to the introduction of automation tools, human engineers need to invest more focused effort during expert committee reviews in the existing process to identify risks and discover issues.

V. EVALUATION

*A. Evaluation Criteria*

The goal of chip verification is to ensure the correctness and completeness of chip functionality implementation. Therefore, accuracy-related metrics are prioritized as the primary evaluation criteria. MAVF's work encompasses the entire process from reading design specifications to generating testbench frameworks. As shown in the framework implementation description, different agents are distinguished, with each agent performing work across multiple stages. Therefore, we evaluate the stage-wise results of the framework's main work. This study uses module-level verification work completed by human engineers as the evaluation baseline. These baseline works include complete verification processes with deliverables that have been thoroughly verified and checked. The evaluation focuses on accuracy, efficiency, and resource dimensions.

Specific accuracy evaluation criteria are as follows:
1. Specification Parsing: Assess whether information extracted from design specification documents has omissions or errors, measuring the error rate as a percentage of incorrect parts against total work
2. Verification Plan: Using manually decomposed test points and test cases as baseline, measure error rates of framework-automated test points and test cases decomposition as a percentage of missing or incorrect items against total test points or test cases
3. Testbench Specification: Evaluate framework, component, and scenario error rates based on the proportion of words requiring human engineer modification against total auto-generated words
4. Code Generation: Evaluate framework, component, and scenario error rates based on the proportion of code lines requiring human engineer modification against total auto-generated lines

Since the total processing time of the full-flow framework for a design document ranges between 40-100 minutes, significantly shorter than manual completion time, efficiency evaluation criteria use the time required for complete manual execution to quantify the effectiveness of this automation framework. In semi-automated framework execution scenarios with human intervention, the time invested by human engineers in the process is quantified.

For resource evaluation, depending on the models used, total costs are calculated by multiplying the actual consumed tokens by their unit prices, thus quantifying the resources consumed by the automation framework execution.

TABLE II
The prices of different models used in the evaluation process

| Name | Input token | Output token |
| --- | --- | --- |
| openai/4o-mini | $0.15/M tokesn | $0.6/M tokens |
| anthropic/claude-3.5-sonnet | $3/M tokens | $15/M tokens |
| deepseek/deepseek-r1 | $0.55/M tokens | $2.19/M tokens |

*B. Evaluation Set*

Due to the current lack of standardized benchmark evaluation sets in this field, to more realistically evaluate the framework's performance, three module-level verification works of different complexity and design scales from actual projects that have completed tape-out verification were used for evaluating various stages of the framework:

TABLE III
Different DUTs (Devices Under Test) used in the evaluation process

| Module Name | Code Size (Lines) | Documentation (Words) | Functionality Description |
| --- | --- | --- | --- |
| MODULE_A | 1706 | 1500 | Support address remapping for multiple address ranges. |
| MODULE_B | 4565 | 5500 | Supports multi-channel DMA with Register and Command list modes |
| MODULE_C | 20495 | 21000 | Supports protocol conversion and multi-Ring management |

*C. Comparative Study*

Through evaluating MAVF in verification processes across different modules, we obtained insightful results, particularly in terms of accuracy and efficiency.

*1)* Accuracy

*a)* Framework outperforms simple prompting

MAVF achieves significantly higher accuracy rates across all stages from documentation writing to code generation compared to simply querying GenAI models without the framework. A summary is provided in Figure 5, while detailed results can be found in Appendix A.1.

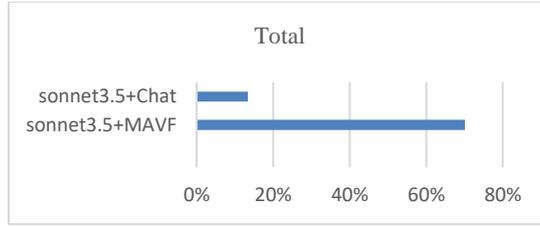

Figure 5. Evaluation Results of Two Different Mode (Summary)
sonnet3.5+MAVF represents tests using anthropic/claude-3.5-sonnet3.5 model with full MODULE_B design specifications as input, running MAVF fully automatically without human intervention. sonnet3.5+Chat represents tests using the same model in conversational mode with full MODULE_B design specification documents as context prompts plus specific task requirements, without human intervention.

The results demonstrate that fully automated MAVF execution achieves significantly higher achievement rates and accuracy across key objectives compared to simply asking GenAI models to complete specified tasks.

*b)* Higher model performance correlates with better framework results

When executing the automated framework process on the same module using different models, we observed a positive correlation between model performance and framework execution results. A summary is provided in Figure 6, while detailed results can be found in Appendix A.2.

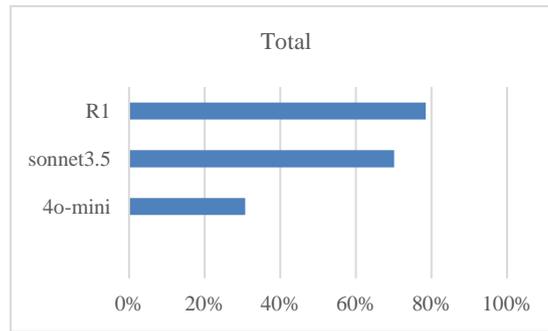

Figure 6. Evaluation Results of Three Different Models (Summary)
4o-mini represents tests using openai/4o-mini model, r1 represents tests using deepseek/deepseek-r1 model, and sonnet represents tests using anthropic/claude-3.5-sonnet3.5 model, all with full MODULE_B design specifications as input and fully automated MAVF execution without human intervention.

The results show that MAVF execution with r1 achieves higher accuracy across metrics compared to both sonnet3.5 and 4o-mini, which aligns with the performance evaluation of these two models. At the same time, it can be observed that the R1 model is more proficient in analysis and specification formulation, while sonnet3.5 is more adept at code generation, which is consistent with the specific characteristics of these two models in their training and inference processes.

*c)* Framework performance decreases with design complexity

When executing the automated process on modules of varying complexity using the same model, we observed a clear negative correlation between design complexity and framework execution accuracy. A summary is provided in Figure 7, while detailed results can be found in Appendix A.3.

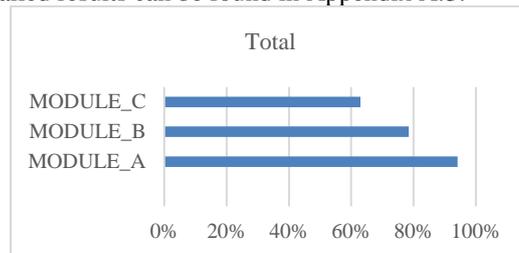

Figure 7. Evaluation Results of Three Different Modules (Summary)
Results show tests using deepseek/deepseek-r1 model on MODULE_A, MODULE_B, and MODULE_C modules respectively, using their full design specifications with fully automated MAVF execution without human intervention.

The results demonstrate that MAVF automation accuracy decreases significantly as design specification complexity increases. Particularly in code generation, the accuracy drops noticeably without human intervention, which also indicates a greater need to strengthen capabilities in the area of code generation.

*2)* Efficiency

In practical applications, since MAVF execution still contains errors, introducing human feedback mechanisms where engineers review and correct MAVF outputs at key stages allows quickly achieving goals that would take human engineers much longer to accomplish alone. A summary is provided in Figure 8, while detailed results can be found in Appendix A.4.

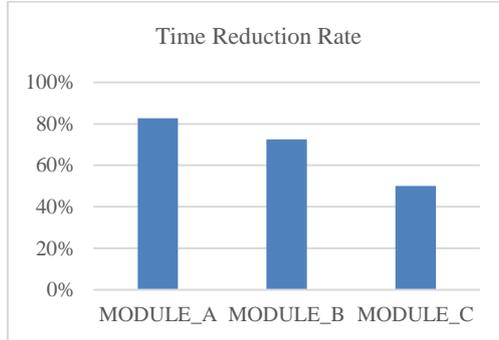

Figure 8 Time reduction rate

Time reduction rate shows the percentage of time saved through MAVF assistance ((human time - human&MAVF time)/human time ×100%) (Summary)

Results show MAVF reduces engineer time investment by 83% for simple modules and 73% for medium-scale modules. There's also a trend that MAVF benefits decrease as design complexity increases.

*3)* Cost

Based on current large model market prices, the resource costs for completing the entire framework are very low, as shown in Table IV.

Table IV

"input" shows data volume sent to models as prompts. "output" shows information volume returned by models to MAVF. "total" shows the cost calculated based on current model prices for total tokens consumed.

| model | MODULE_A | | | MODULE_B | | | MODULE_C | | |
| --- | --- | --- | --- | --- | --- | --- | --- | --- | --- |
| | input | output | total | input | output | total | input | output | total |
| 4o-mini | 378k | 23k | $0.07 | 775k | 23k | $0.13 | 878k | 47k | $0.16 |
| sonnet3.5 | 402k | 33k | $1.69 | 603k | 50k | $2.55 | 1204k | 79k | $4.79 |
| r1 | 545k | 49k | $0.20 | 807k | 76k | $0.30 | 1080k | 111k | $0.84 |

This demonstrates that using MAVF to assist chip verification work offers excellent cost-effectiveness, achieving substantial benefits with minimal resource investment. Here we can see that the R1 model's output token count is significantly higher than sonnet3.5, which is due to the chain-of-thought reasoning in the r1 model requiring more output tokens.

## VI. DISCUSSION

Based on the evaluation results, MAVF shows significant improvements compared to conventional dialogue-based approaches in generating verification documents or testbench code. While the framework's performance decreases when dealing with complex designs, it demonstrates better capabilities when using higher-performance models for more complex design scenarios.

The evaluation process reveals that when human intervention is introduced at key stages to optimize the framework's output for complex designs, there is a substantial efficiency improvement compared to purely manual completion of related tasks. From a resource perspective, it can be concluded that the resource costs required are negligible compared to the efficiency gains achieved.

Although the evaluation results show significant effectiveness, there is currently a lack of standardized evaluation sets to help conduct more reliable performance tuning. More work needs to be done on the framework's adaptability. For instance, different modules vary greatly in their functional characteristics. Future work could involve categorizing all possible types of module characteristics and conducting optimization evaluations for different types.

The effectiveness of the MAVF framework is primarily manifested in three aspects: models, multi-agent framework, and domain workflow. Regarding models, with the development of pre-trained models and chain-of-thought reasoning models, the foundational model capabilities are increasingly improving. Combined with industry data for supervised learning or fine-tuning, the models' inherent capabilities will better meet chip verification scenarios. For the framework aspect, there is considerable room for improvement in reliability, maintainability, and user-friendliness for human engineer interaction while addressing various practical issues in the engineering process. Additionally, better methods can certainly be implemented in quality assurance mechanisms. In terms of domain workflow, as our understanding of GenAI model capabilities becomes more accurate, we can decompose chip verification work into decoupled steps and phases that better match model capabilities, enabling more clever and elegant applications of GenAI models in chip verification processes.

This research not only provides an innovative solution for chip verification automation but also offers valuable practical examples for applying large language models in professional domains. The research outcomes have significant theoretical and practical implications for advancing the intelligent development of integrated circuit design.

## Appendix A. EVALUATION RESULT DETAILS

*A.1 Evaluation Results of Two Different Mode*

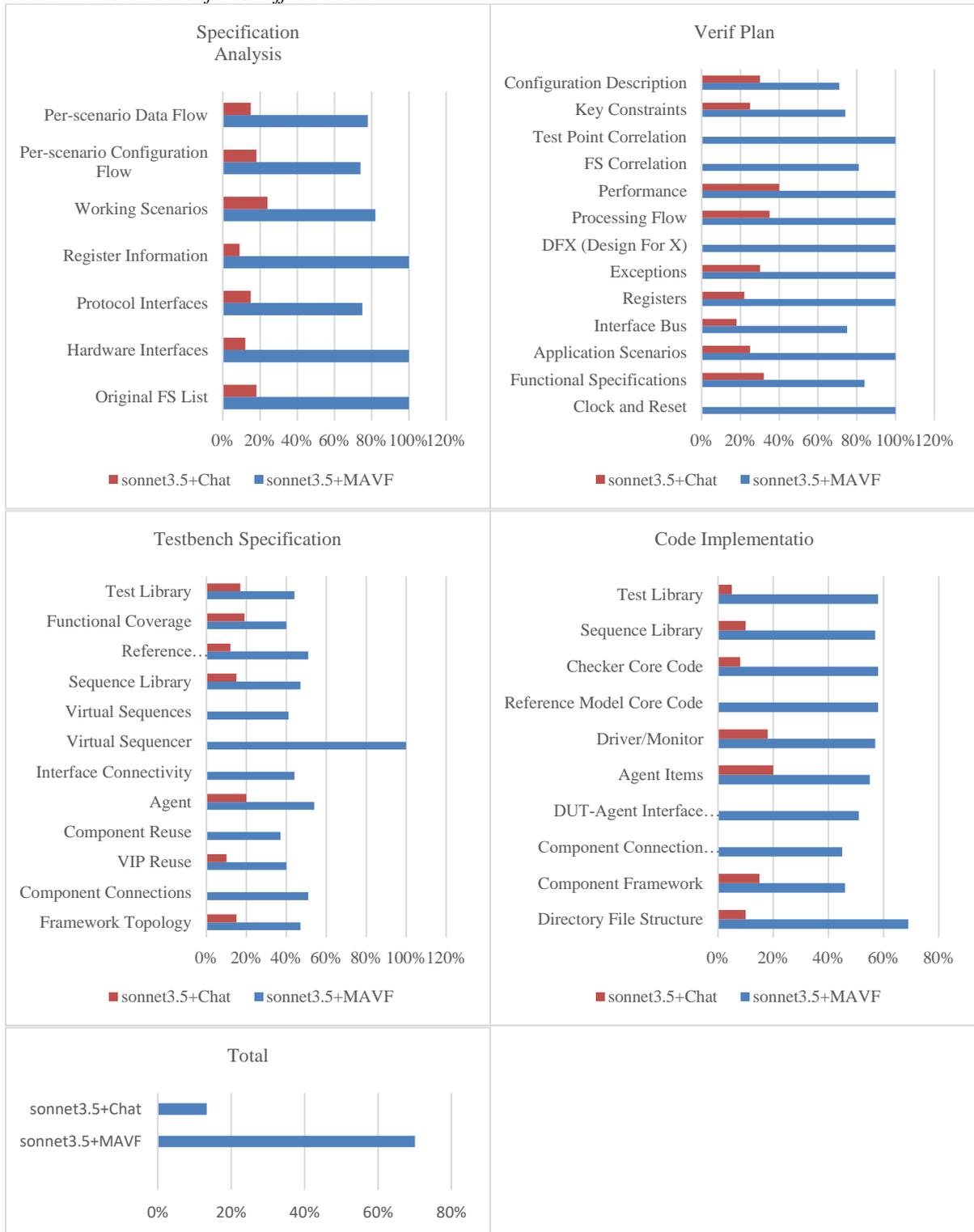

Figure 9. Evaluation Results of Two Different Mode (Details)

sonnet3.5+MAVF represents tests using anthropic/claude-3.5-sonnet3.5 model with full MODULE_B design specifications as input, running MAVF fully automatically without human intervention. sonnet3.5+Chat represents tests using the same model in conversational mode with full MODULE_B design specification documents as context prompts plus specific task requirements, without human intervention.

*A.2 Evaluation Results of Three Different Models*

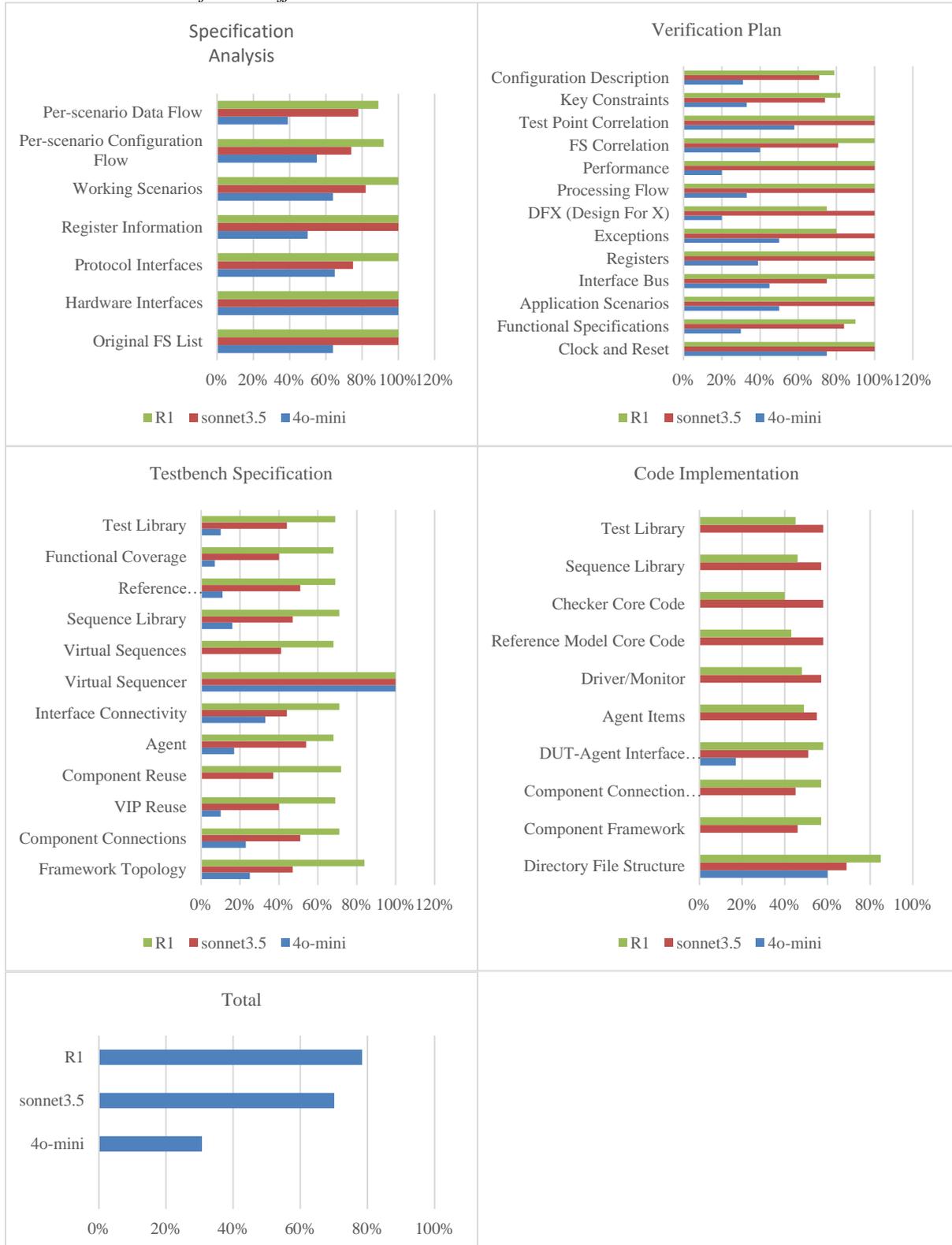

Figure 10. Evaluation Results of Three Different Models (Details)
4o-mini represents tests using openai/4o-mini model, r1 represents tests using deepseek/deepseek-r1 model, and sonnet represents tests using anthropic/claude-3.5-sonnet3.5 model, all with full MODULE_B design specifications as input and fully automated MAVF execution without human intervention.

*A.3 Evaluation Results of Three Different Modules*

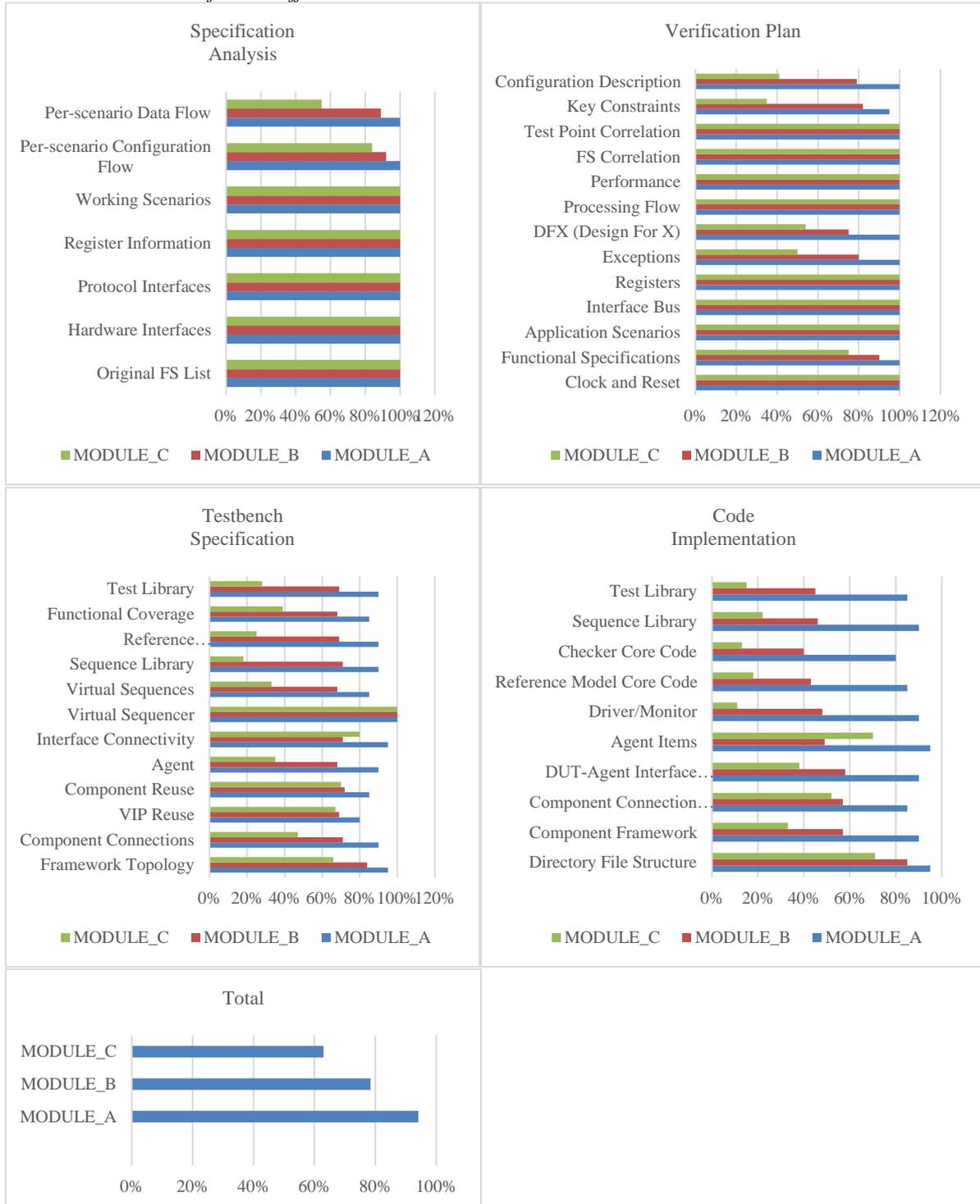

Figure 11. Evaluation Results of Three Different Modules (Details)
Results show tests using deepseek/deepseek-r1 model on MODULE_A, MODULE_B, and MODULE_C modules respectively, using their full design specifications with fully automated MAVF execution without human intervention.

*A.4 Time requirements for MAVF mode versus human mode*

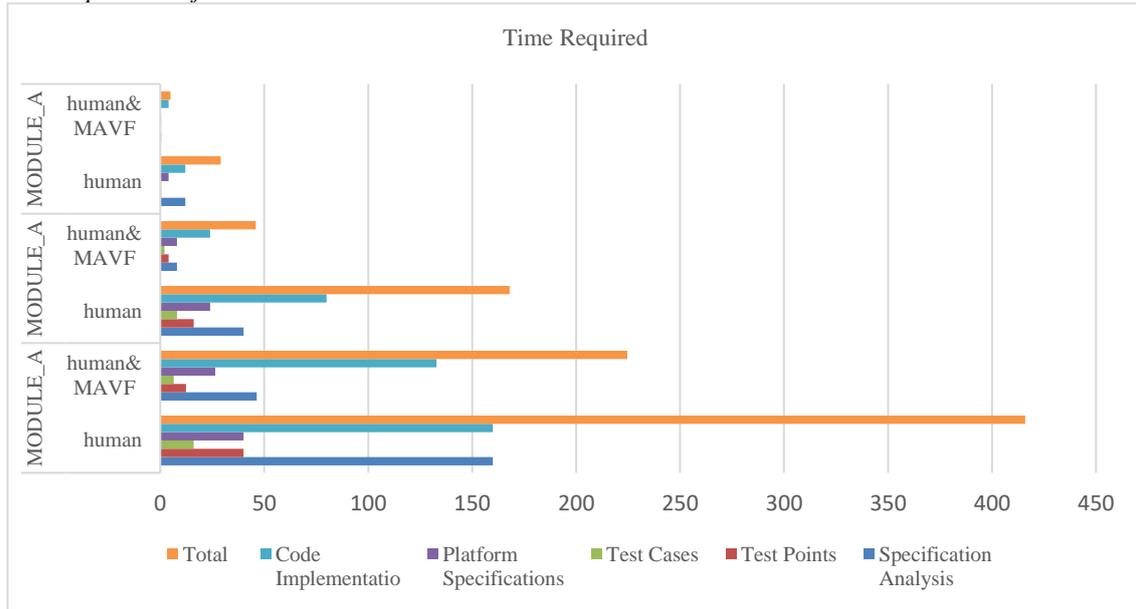

Figure 12. Time requirements (Details)

"human" shows time (in hours) required for human engineers to complete tasks independently. "human&MAVF" shows time (in hours) required to meet the same delivery standards with MAVF assistance.

# Appendix B. PARTIAL EXECUTION EFFECT DEMONSTRATION

*B 1. Comparison of the effectiveness*

Table V

Comparison of the effectiveness of Module_A in two scenarios: fully automated execution in MAVF versus test point generation through conversational interaction with LLM. Due to space limitations, only partial content is presented.

| # | MAVF | # | Chat |
|---|---|---|---|
| 1 | `"TP": [` | 1 | 1. Address expansion functionality verification: |
| 2 | `  {` | 2 | (a) Verify that M7 (32-bit address bus) can correctly access N2 (48-bit address bus) address space through jm_soc_remap |
| 3 | `    "tp_l1_name": "Address Width Conversion Function",` | 3 | (b) Test address translation from 32-bit to 48-bit addresses |
| 4 | `    "tp_l2": [` | 4 | 2. Window configuration verification: |
| 5 | `      {` | 5 | (a) Test different window quantity configurations (16, 32, 64) |
| 6 | `        "tp_l2_name": "Address Width Parameter Test",` | 6 | (b) Verify start and end address configuration for each window |
| 7 | `        "tp_l3": [` | 7 | (c) Test window enable/disable functionality (default disabled) |
| 8 | `          " Supports S_ADDR_W bit configuration",` | 8 | (d) Verify 1MB minimum granularity for window addresses |
| 9 | `          " Supports M_ADDR_W bit configuration"` | | |
| 10 | `        ]` | | |
| 11 | `      },` | | |
| 12 | `      {` | | |
| 13 | `        "tp_l2_name": "Address Conversion Calculation Test",` | | |
| 14 | `        "tp_l3": [` | | |
| 15 | `          "When window matches, output address = {ext_addr, 20'd0} + axi_slave_addr - region_start_addr",` | | |
| 16 | `          "When no window matches, output address = {default_slave_address[M_ADDR_W-1:12], axi_slave_addr[11:0]}"` | | |
| 17 | `        ]` | | |
| 18 | `      }` | | |
| 19 | `    ]` | | |
| 20 | `  },` | | |
| 21 | `  {` | | |
| 22 | `    "tp_l1_name": "Window Configuration Function",` | | |
| 23 | `    "tp_l2": [` | | |
| 24 | `      {` | | |
| 25 | `        "tp_l2_name": "Window Quantity Configuration Test",` | | |
| 26 | `        "tp_l3": [` | | |
| 27 | `          "Supports configuration of 1-64 windows",` | | |
| 28 | `          "Register writes beyond the configured window count are invalid and read as 0"` | | |
| 29 | `        ]` | | |
| 30 | `      },` | | |
| 31 | `      {` | | |
| 32 | `        "tp_l2_name": "Window Address Configuration Test",` | | |
| 33 | `        "tp_l3": [` | | |
| 34 | `          "Start address and end address minimum granularity is 1MB",` | | |
| 35 | `          "Addresses must be 8-byte aligned"` | | |
| 36 | `        ]` | | |
| 37 | `      }` | | |
| 38 | `    ]` | | |
| 39 | `  }` | | |
| 40 | `]` | | |

*B 2. Automatically Generated Testbench Top Topology for MAVF*

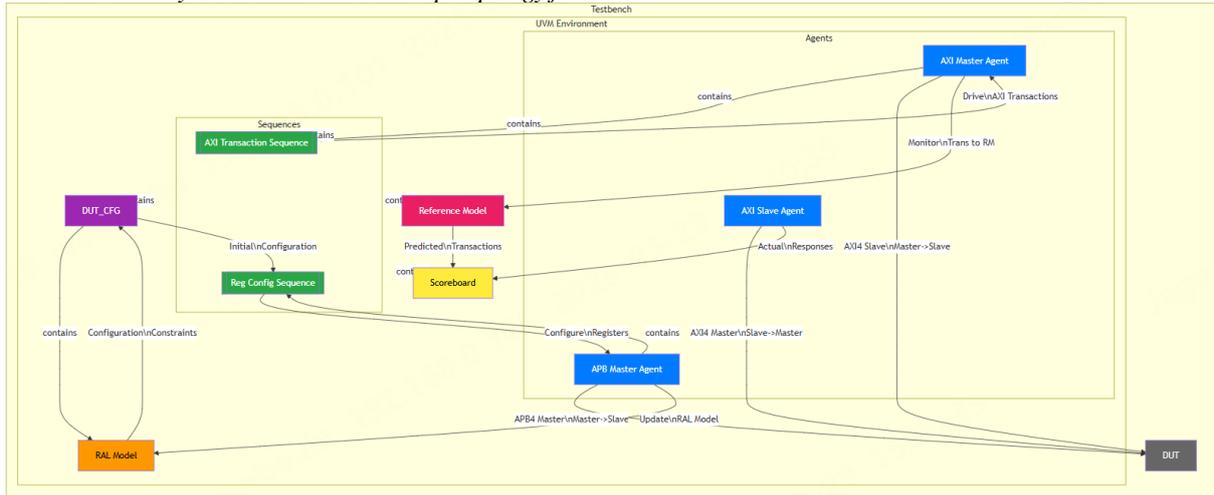

Figure 13
Topology of the testbench generated entirely by MAVF without human intervention according to MODULE_A specifications. This visualization is also automatically generated by MAVF based on the Mermaid language.